\documentclass[onecolumn,preprintnumbers,elsart]{revtex4}

\usepackage{amsmath}
\usepackage{cases}
\usepackage{mathrsfs}
\usepackage{graphicx}
\usepackage{dcolumn}
\usepackage{bm}
\usepackage[center]{subfigure}
\usepackage{color}

\begin{document}

\title{Discrete solitons in waveguide arrays with long-range linearly coupled effect}

\author{Zhijie Mai$^{1,2}$}
\email{zhijiemai@gmail.com}
\author{Shenhe Fu$^{1}$}
\author{Jianxiong Wu$^{3}$}
\author{Yongyao Li$^{2}$}

\affiliation{$^{1}$State Key Laboratory of Optoelectronic Materials and Technologies,\\
School of Physics and Engineering, Sun Yat-sen University, Guangzhou 510275, China\\
$^{2}$Department of Applied Physics, South China Agricultural University, Guangzhou 510642, China\\
$^{3}$Department of Electrical and Computer Engineering, University of Toronto, Toronto, M5S3G4, Canada.
}

\begin{abstract}
We study the influences to the discrete soliton (DS) by introducing linearly long-range nonlocal interactions, which give rise to the off-diagonal elements of the linearly coupled matrix in the discrete nonlinear schrodinger equation to be filled by non-zero terms. Theoretical analysis and numerical simulations find that the DS under this circumstance can exhibit strong digital effects: the fundamental DS is a narrow one, which occupies nearly only one waveguide, the dipole and double-monopole solitons, which occupy two waveguides, can be found in self-focusing and -defocusing nonlinearities, respectively. Stable flat-top solitons and their stagger counterparts, which occupy a controllable number of waveguides, can also be obtained through this system. Such digital properties may give rise to additional data processing applications and have potential in fabricating digital optical devices in all-optical networks.\\


\end{abstract}

\maketitle
\section{Introduction}

Nonlinear discrete systems exhibit many physical characteristics and attract great attention in many branches of physics \cite{Lederer}. The evolution of nonlinear waves inside discrete systems is a popular topic in the subjects of optics \cite{Christodoulides} and Bose-Einstein condensates (BEC) \cite{Trombettoni}.

In optics, the basic model of a nonlinear discrete system is an array of evanescently coupled waveguides made by nonlinear materials. Light propagation in waveguide arrays is primarily characterized by the coupling caused by the overlap between the fundamental modes of the nearest neighboring waveguides. One can then effectively model the power exchanging process in the array according to coupled-mode theory, also known as tight-binding approximation, and can give rise to a discrete nonlinear Schrodinger equation \cite{Christodoulides2}:
\begin{eqnarray}
i{\partial\over\partial z}u_{n}=-C_{0}(u_{n-1}+u_{n+1})+\gamma|u_{n}|^{2}u_{n}.\label{DNLS}
\end{eqnarray}
where $u_{n}$ is the field amplitude of the $n$-th mode. Because each waveguide is identical, for simplification, the propagation constant $\beta$ is absorbed into the phase of $u_{n}$. $z$ is the propagation distance along the waveguides, $n$ is the waveguide number, $\gamma$ is the nonlinear parameter of the waveguide, and the coefficient $C_{0}$ defines the coupling which depends on the optical wavelength and the field overlap between the neighboring waveguides. In Eq. (\ref{DNLS}), if we use conversion $z\rightarrow t$ , the equation can be used to describe the BEC trapped in a deep optical lattice after mean-field approximation \cite{Gligoric}.  $u_{n}$ is the strength of the strongly localized mode in the vicinity of the lattice site. $\gamma$ is the contact nonlinearity induced by Feshbach resonance \cite{Inouye}. Generally, Eq. (\ref{DNLS}) can be expressed as the matrix form
\begin{eqnarray}
i{\partial\over\partial z}U=(C+V)U, \label{DNSL_matrix}
\end{eqnarray}
where, the matrix $U$ and the elements of matrix $V$ are defined as $U=\left(u_{1},\cdots,u_{N}\right)^{T}$ (the superscript T indicates the transposition of the matrix, $N$ is the number of waveguides) and $V_{mn}=\gamma|u_{m}|^{2}\delta_{mn}$ ($\delta_{mn}$ is the Kronecker symbol), respectively.
The neighboring linear interaction gives rise to the elements of the $C$ matrix (i.e., the linearly coupled matrix) as a kind of tridiagonal matrix
\begin{eqnarray}
C_{mn}=-C_{0}\left(\delta_{m,n-1}+\delta_{m,n+1}\right)
\end{eqnarray}

In nonlinear systems, a crucial issue is to study the formation and properties of discrete solitons (DSs). DS formation can be intuitively understood as a balance between on-site nonlinearity and discrete diffraction induced by linear coupling among adjacent waveguides or lattice sites. DSs provide strong potential in all-optical data processing applications; more specifically, they can realize intelligent functional operations, such as routing, blocking, logic functions and time-gating, in many all-optical devices \cite{Christodoulides3}.

However, an interesting extension is to investigate the formation of DS encompassing a long-range linearly coupled effect. In \cite{Kevrekidisa}, Kevrekidis and his co-workers study the properties of DSs in an NNN (next-nearest neighbor) model, in which the linearly coupled matrix becomes a quadruple-diagonal matrix after higher-order diffraction is considered. Recently, significant studies of the nonlinear dipolar field in the nanoparticle train by Noskov and his co-workers \cite{Noskov}, which can be viewed as a discrete nonlinear system, report that the linearly coupled effect can exist among the entire lattice sites because of long-range dipole-dipole interaction. This system can produce all the non-zero off-diagonal elements in the linearly coupled matrix.

In this letter, we consider the formation of DSs after employing the non-zero off-diagonal elements in the coupled matrix. After such a cross-coupling effect is introduced to the waveguide array, the family of solitons exert many digital effects, occupy a controllable number of waveguides by means of power, and can thus be considered digital DSs. Such digital information (the number of occupied waveguides) offers the soliton signal additional data processing ability.

We give a brief description to the model in section II and carry out the numerical simulations in section III. It is interesting to find that the ground state discrete soliton has a super narrow width, which occupies nearly only one waveguide. While for the excited state solitons, stable double monopole solitons, dipole solitons and flat-top solitons are found in the case of self-focusing or -defocusing. The paper is concluded in Sec. IV.
\section{Model}
\begin{figure}[tbp]
\centering%
\subfigure[] {\label{fig11a}
\includegraphics[scale=0.4]{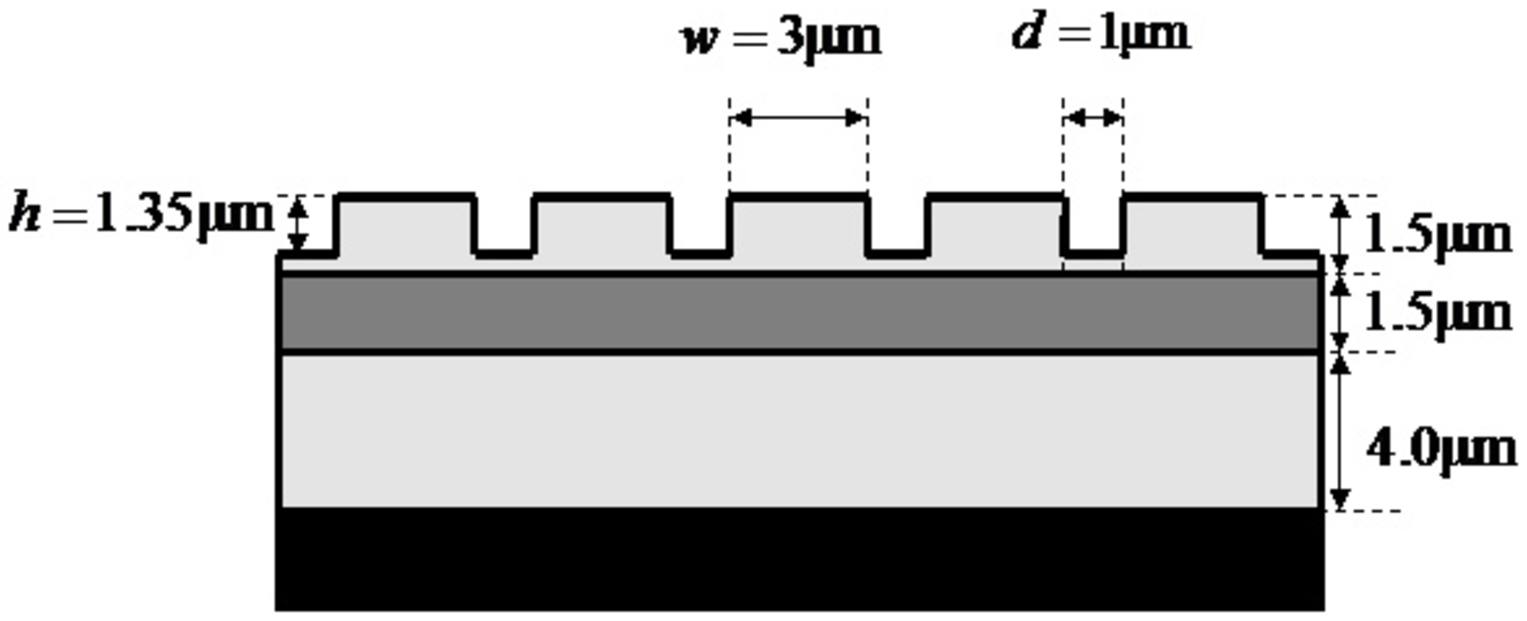}}%
\subfigure[] {\label{fig11b}
\includegraphics[scale=0.4]{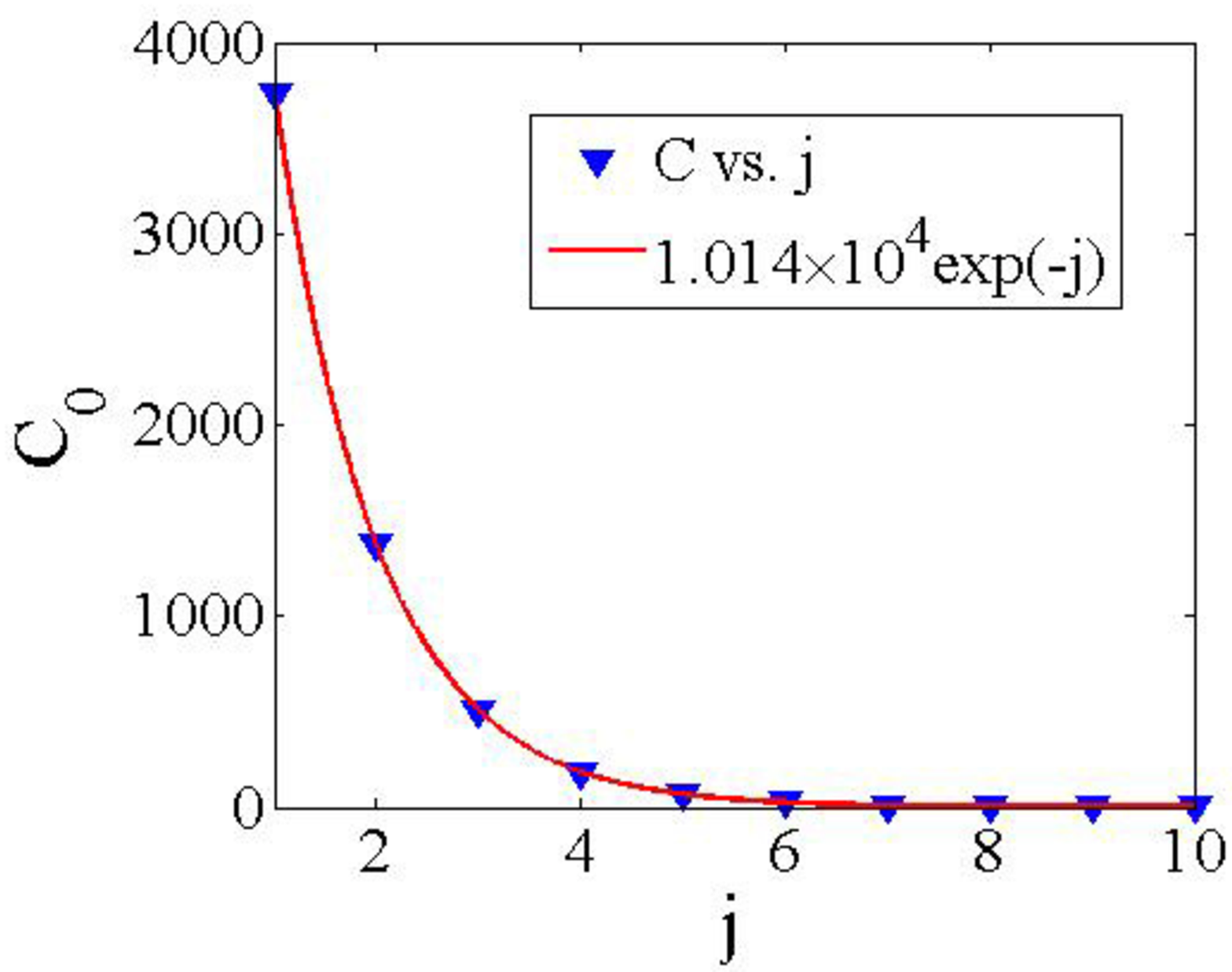}}
\caption{(Color online) (a) Experiment scheme of waveguide array. The light gray color areas are the Al$_{0.24}$Ga$_{0.76}$As, the dark gray color area is the Al$_{0.18}$Ga$_{0.82}$As, and the black color area is the GaAs substrate. (b)Linearly coupled parameter versus $j$, which can be fitted by a function of $A\exp(-j)$.}\label{Fundsolution}
\end{figure}

In the model, we consider the matrix elements of $C$ in Eq. (\ref{DNSL_matrix}) given by
\begin{eqnarray}
C_{mn}=
\begin{cases}
c_{0}\exp(-j) & (j\neq 0) \\
0 & (j=0)
\end{cases} \label{C_element}
\end{eqnarray}
where $c_{0}$ is the controlled parameter and $j=|m-n|$. Eq. (\ref{C_element}) forms the linearly coupled matrix $C$ exhibiting long-range nonlocal linearly coupled interactions among each lattice site. In optics, when the separation between waveguides is narrow enough, such higher-order cross coupling effect can be induced \cite{GordonR}. According to this definition, Eq. (\ref{DNLS}) is revised as
\begin{eqnarray}
i{\partial\over\partial z}u_{n}=\gamma|u_{n}|^{2}u_{n}-\sum_{m}C_{mn}u_{m}. \label{Nonlocal_DNSL}
\end{eqnarray}
where $\gamma$ is the fixed nonlinear parameter of the system and $\gamma=1$ and $\gamma=-1$ indicate the system features of self-focusing and -defocusing nonlinearity, respectively. The total power (in the system of BEC, it should be replaced by `total norm') of the guide mode in the system is given as $P=\sum^{N}_{n}|u_{n}|^{2}$.

In optics, we can design an experiment scheme to apply our model. We assume the model is built based on an AlGaAs single-mode waveguide structure modified from \cite{Joushaghani}, with a real scale displayed in Fig. \ref{fig11a}. Varied composition of aluminum in each layer provides the refractive index difference. Given that the fundamental mode is confined vertically in the Al$_{0.18}$Ga$_{0.82}$As layer and laterally between the walls of the ridge structure, the effective index method is valid for collapsing the 2D array into an equivalent 1D structure. In this case, the effective index is 3.263 for the guiding and 3.259 for the background. Fig. \ref{fig11b} shows that the real-scale linearly coupled parameter between 0 and the $j^{th}$ waveguide, which is calculated numerically from the real scale of the experimental scheme depicted in Fig. \ref{fig11a} by means of the coupled mode theory \cite{website}, can be fitted by the function of $A\exp(-j)$ with $A=1.01\times10^{4}$ m$^{-1}$; this fitting can correctly describe the nearest-neighboring and high-order coupling between waveguides. The real-scale nonlinear coefficient of this structure is approximated as $\kappa=$6.5 W$^{-1}$m$^{-1}$ \cite{Aitchison}. The relationship between the scaled parameters in Eq. (\ref{Nonlocal_DNSL}) and these real-scale parameters are given as
\begin{eqnarray}
z=A z',\quad c_{0}={C_{0}\over A},\quad \gamma={P_{0}\kappa\over A},\quad \mathrm{and}\quad u_{n}={a_{n}\over\sqrt{P_{0}}}, \label{Rescaled}
\end{eqnarray}
where $z'$, $C_{0}$ and $a_{0}$ are the real-scale propagation distance, the strength of the linearly coupled parameter, and the guide mode in the waveguide, respectively. $P_{0}$ is the unit of the scaled total power $P$ (the real-scale total power is defined as $P\times P_{0}$). If we fix $|\gamma|=1$, we can obtain $P_{0}\approx1.55$ kW. Because the core area of the waveguide approximates 4.05 $\mu$m$^{2}$, we can estimate that the unit of the power density for the laser source is 0.4 mW/mm$^{2}$, which is reasonable for experimental realization.

We assume that the soliton solutions of Eq. (\ref{Nonlocal_DNSL}) are written as
\begin{eqnarray}
u_{n}(z)=\phi_{n} e^{-i\mu z}, \label{stationary_solution}
\end{eqnarray}
where $\phi_{n}$ is the stationary solution and $-\mu$ is the propagation constant (in the system of BEC, $\mu$ is represented by the chemical potential), which is defined as
\begin{eqnarray}
\mu={U^{\dag}(C+V)U\over P} \label{MU}
\end{eqnarray}

The stability of the stationary mode can be numerically identified by computing the eigenvalues for small perturbations and direct simulations. The perturbed solution is given as $u_{n}=e^{-i\mu z}(\phi_{n}+w_{n}e^{i\lambda z}+v^{\ast}_{n}e^{-i\lambda^{\ast}z})$. Substitution of this solution into Eq. (\ref{DNSL_matrix}) and linearization lead to the eigenvalue problem,
\begin{eqnarray}
\left(
\begin{array}{cc}
C-\mu+2V & \Phi \\
-\Phi^{\ast} & -C+\mu-2V
\end{array}\right)
\left(
\begin{array}{c}
w \\
v
\end{array}\right)
=\lambda\left(
\begin{array}{c}
w\\
v
\end{array}\right), \label{eigenvalue}
\end{eqnarray}
where the elements of the matrix $\Phi$ are defined as $\Phi_{mn}=\gamma\phi^{2}_{m}\delta_{mn}$.
The solution $\phi$ is stable if all the eigenvalues $\lambda$ are real.

\begin{figure}[tbp]
\centering%
\subfigure[] {\label{fig1a}
\includegraphics[scale=0.4]{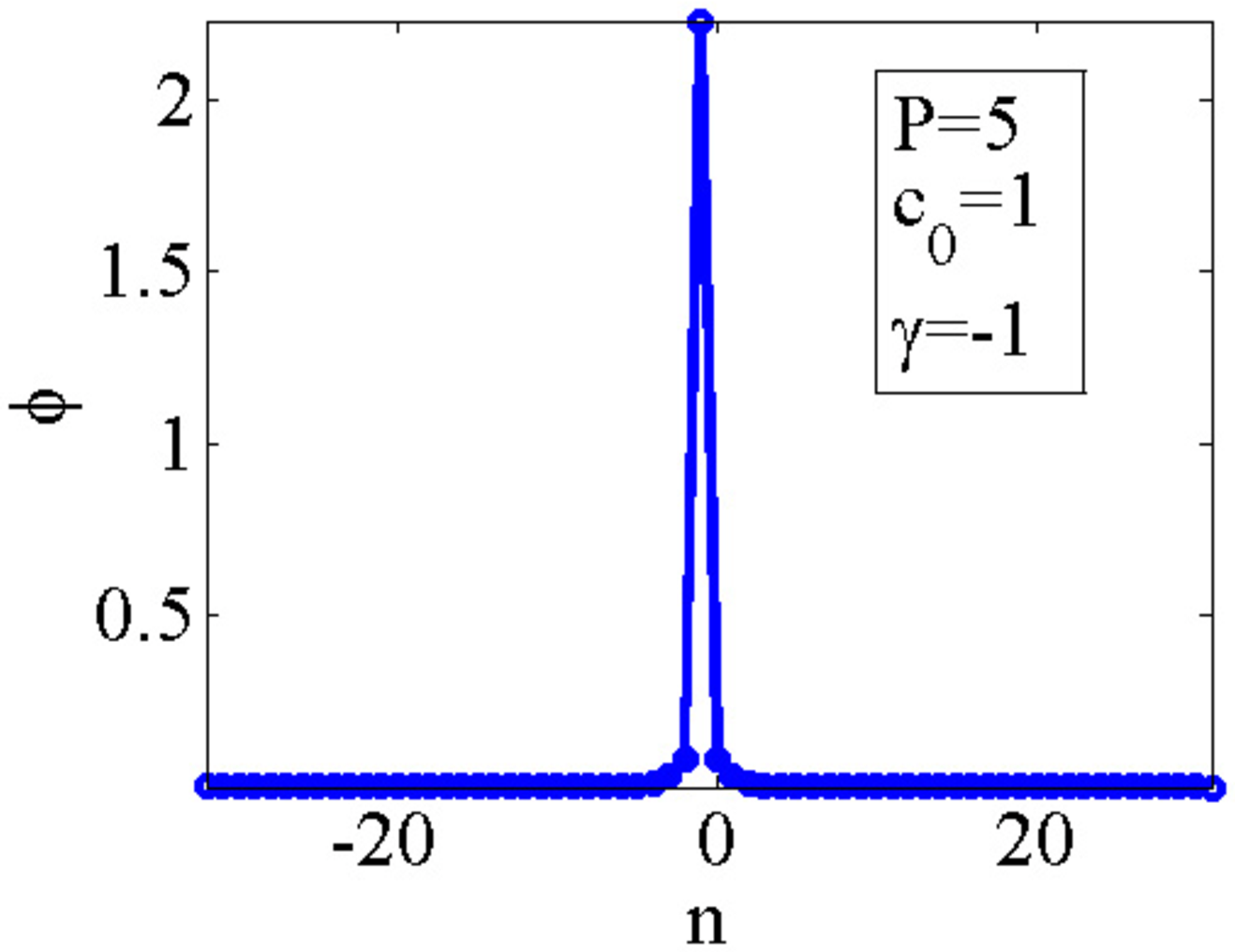}}%
\subfigure[] {\label{fig1c}
\includegraphics[scale=0.4]{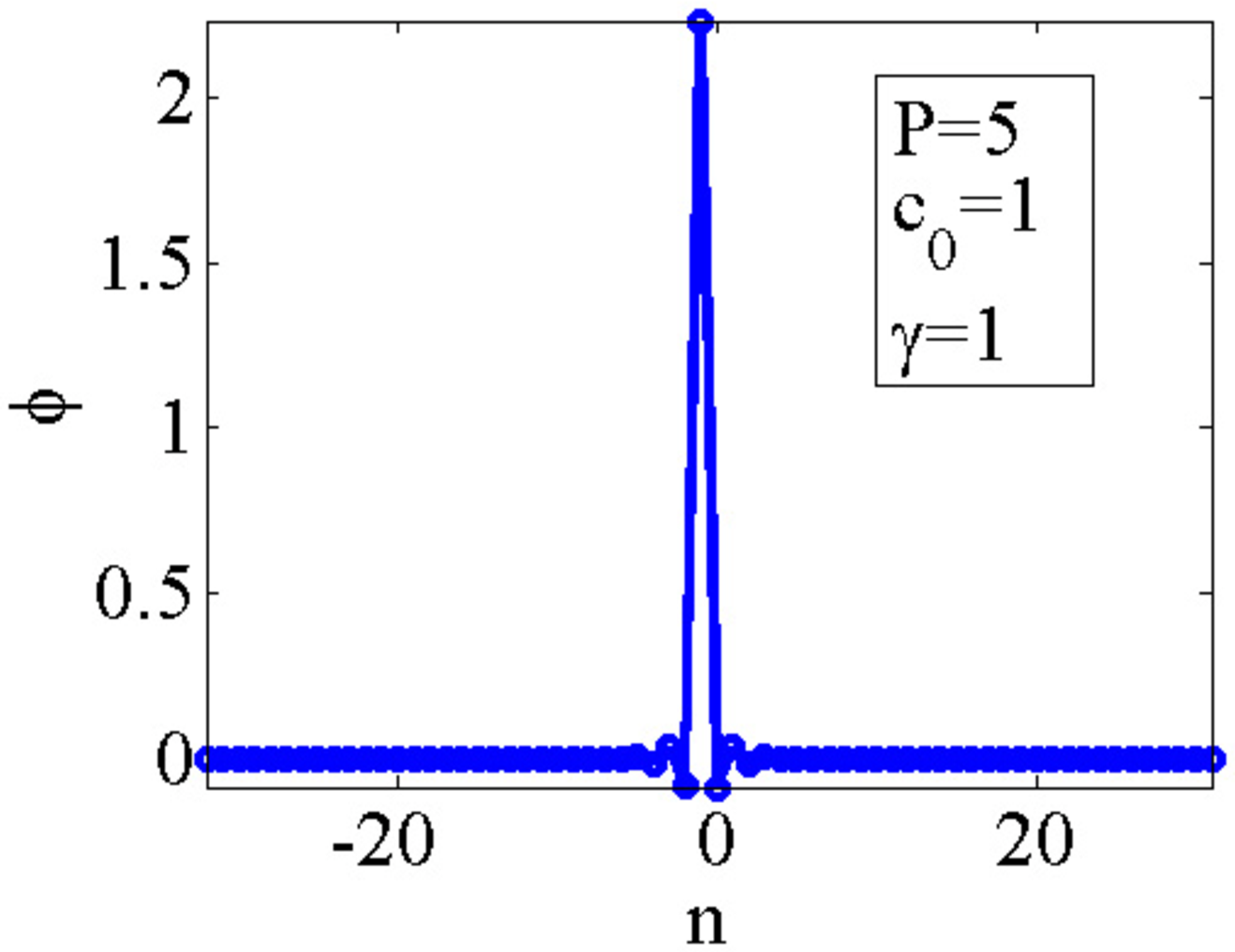}}\\
\subfigure[] {}
\includegraphics[scale=0.4]{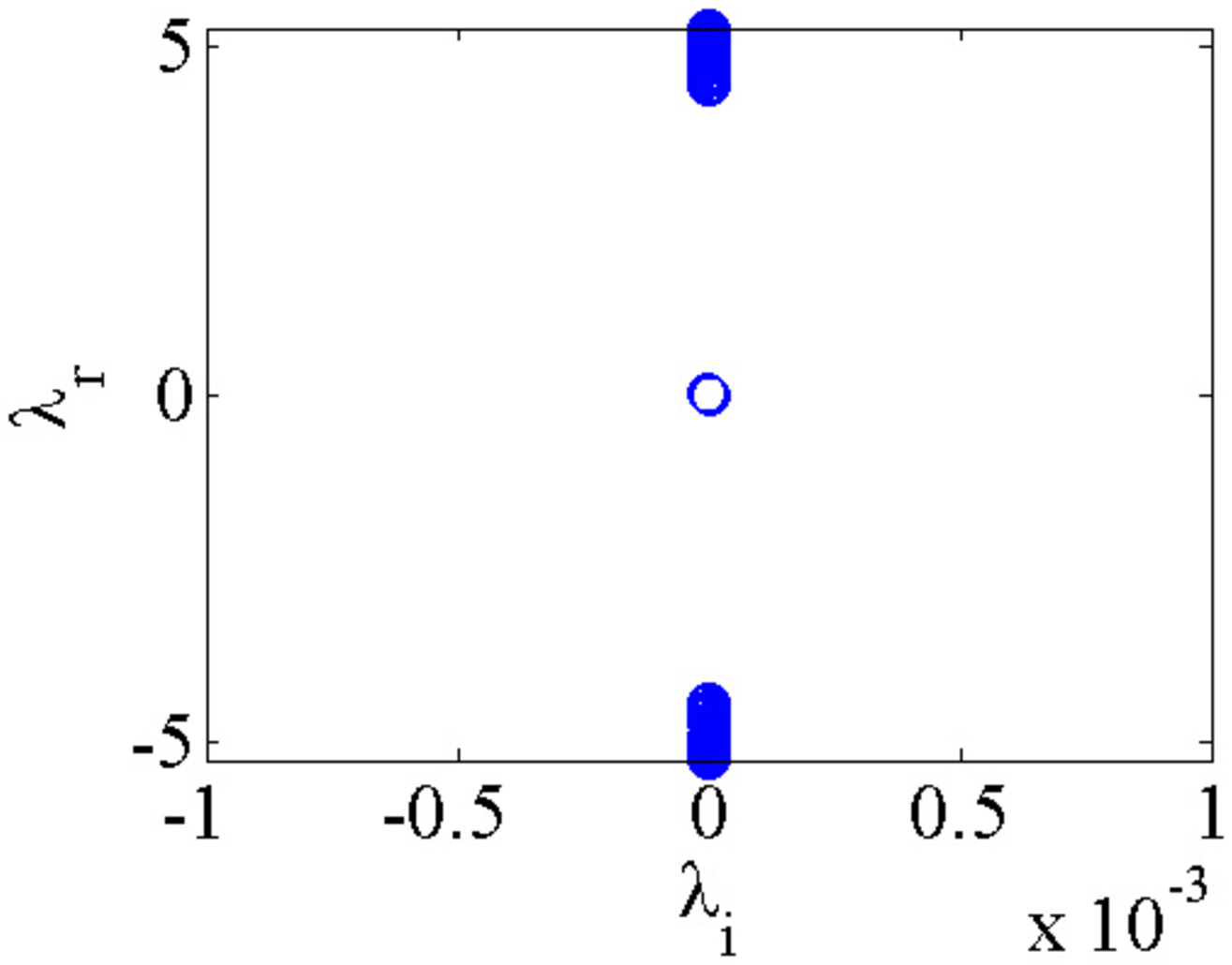}
\subfigure[] {}
\includegraphics[scale=0.4]{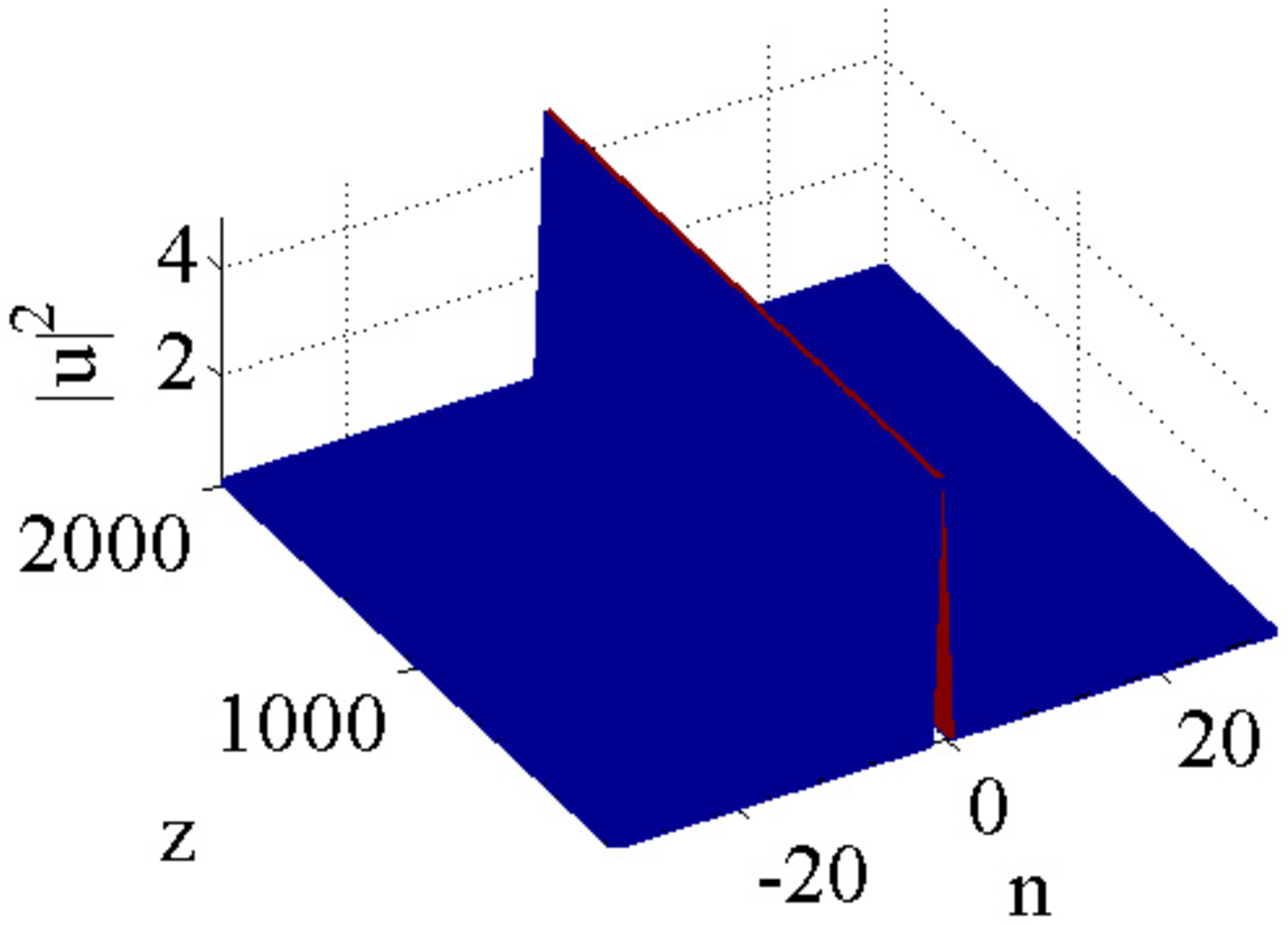}
\caption{(Color online) (a)Fundamental soliton solution in self-attractive/focusing case ($\gamma=-1$) with $c_{0}=1$ and $P=5$. (b) Soliton solution in self-repulsive/defocusing ($\gamma=1$) case, which is staggered from the solution in panel (a). (c) Growth rate of soliton in panel (a), which indicates stability of soliton. (d) Evolution of soliton solution in panel (b), which proves such soliton is also stable.}\label{Fundsolution}
\end{figure}

\begin{figure}[tbp]
\centering%
\subfigure[] {\label{fig2a}
\includegraphics[scale=0.5]{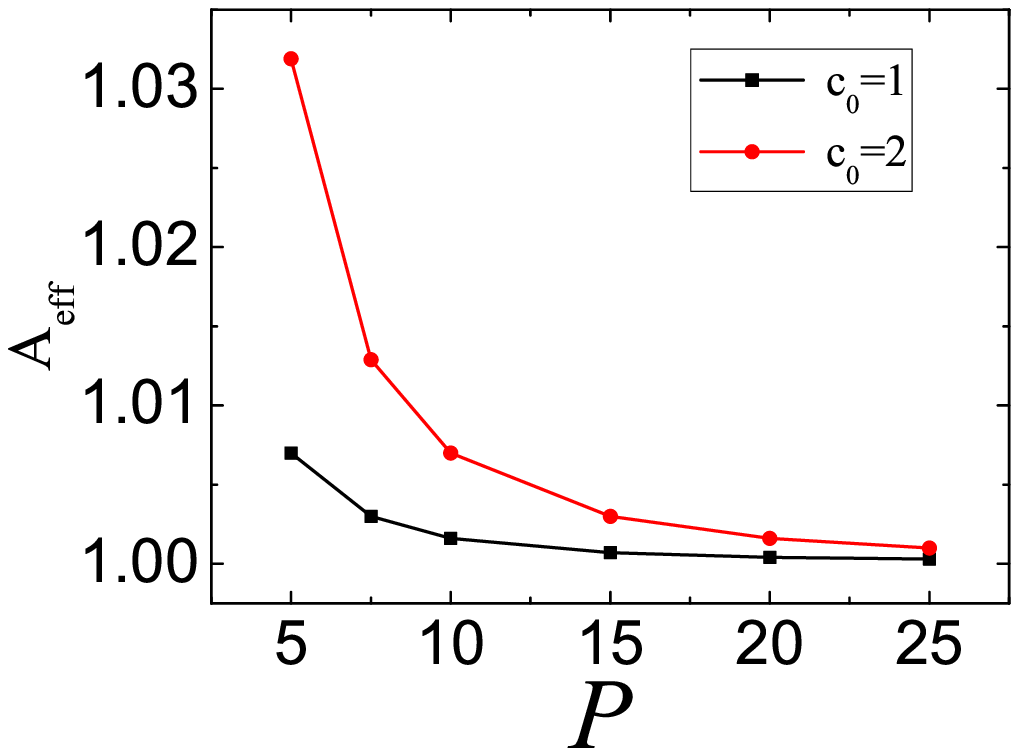}}%
\subfigure[] {\label{fig2b}
\includegraphics[scale=0.5]{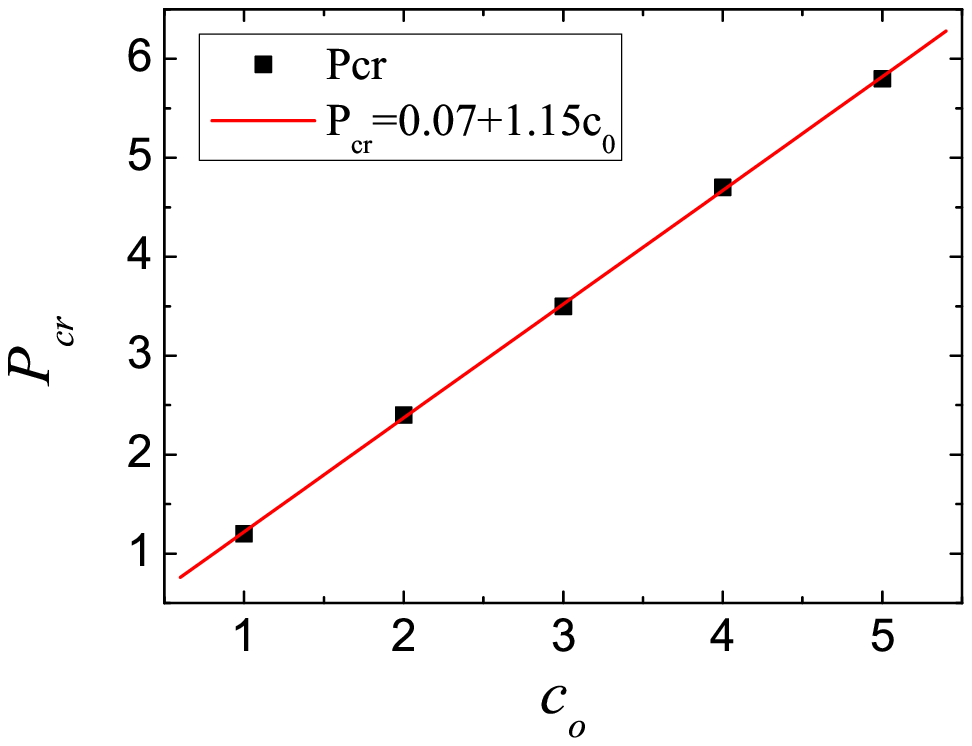}}\\
\subfigure[] {\label{fig2c}
\includegraphics[scale=0.5]{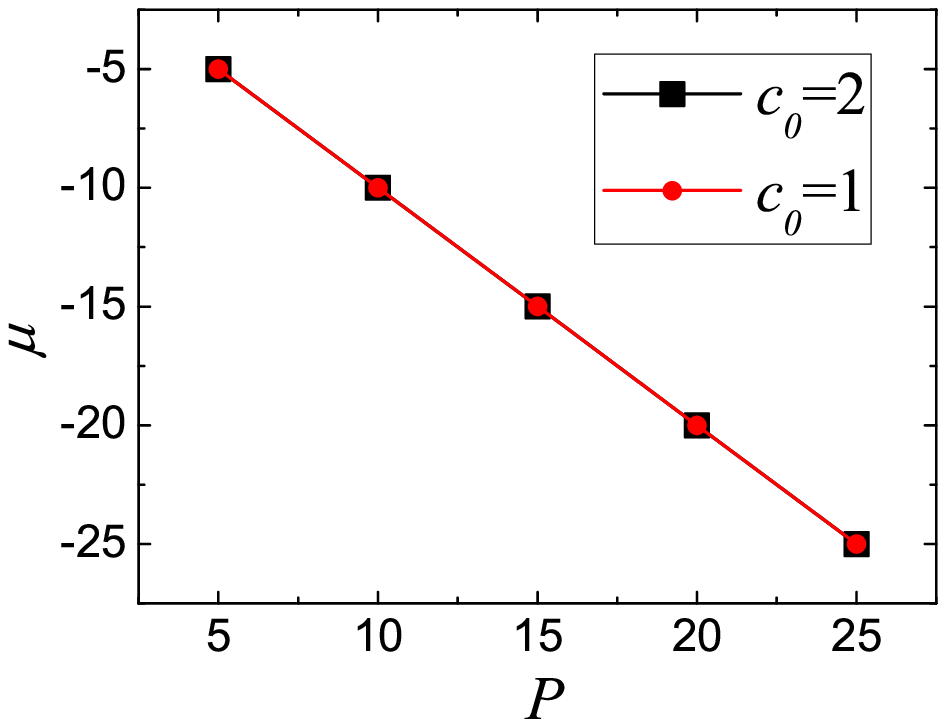}}
\subfigure[] {\label{fig2d}
\includegraphics[scale=0.5]{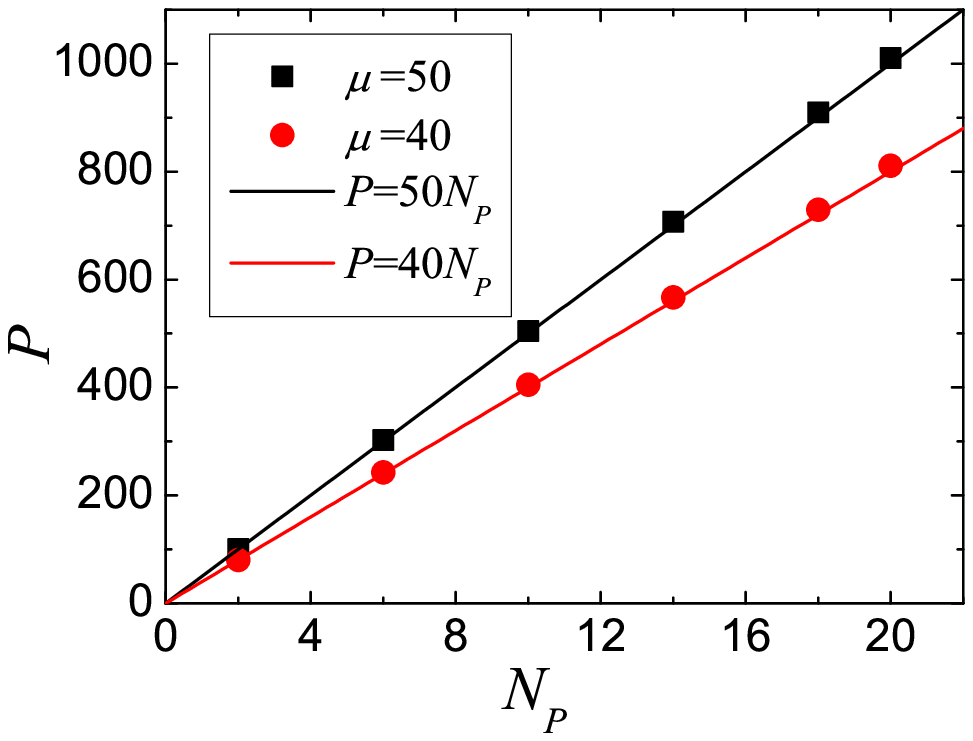}}
\caption{(Color online) (a) Effective area of soliton versus $P$ with different values of $c_{0}$. (b) Critical power as a function of $c_{0}$. (c)$\mu-P$ relation of soliton. (d) Total power of flat-top solitons versus their occupied waveguide numbers with different values of $\mu$. }\label{Fundsolution}
\end{figure}

\section{Numerical Results}
In numerical simulation, we applied the imaginary time propagation (ITP) method  \cite{Chiofalo} to study the fundamental solution in Eq. (\ref{Nonlocal_DNSL}). Figs. \ref{fig1a} gives a typical example of the self-focusing ($\gamma=-1$) case. The effective area, defined as
\begin{eqnarray}
A_{\mathrm{eff}}={\left(\sum^{N}_{n=1}|\phi|^{2}\right)^{2}\over\sum^{N}_{n=1}|\phi|^{4}}, \label{Aeff}
\end{eqnarray}
is shown in Fig. \ref{fig2a}. The effective areas of the soliton are approximated to be equal to 1. The solitons are stable and pinned, and cannot be kicked to move along the lattices. By applying the stagger operation $\phi_{n}'$=$(-1)^{n}\phi_{n}$, we can obtain the stable fundamental gap soliton in the self-defocusing ($\gamma=1$) case. Figs. \ref{fig1c} shows an example of a stable gap soliton staggered from Fig. \ref{fig1a}. Further simulations find that for a fixed value of $c_{0}$, a threshold power value exists to form fundamental solitons. Figure \ref{fig2b} shows the threshold power/norm (namely $P_{\mathrm{cr}}$) as a function of $c_{0}$, indicating the linear growth of $P_{\mathrm{cr}}$ with the linear increase of $c_{0}$. This relation is naturally understood as the higher $c_{0}$, which strongly enhances the linearly coupled effect in all lattice site, and requires stronger nonlinearity, i.e., stronger power, to balance the spreading of the field and form a soliton. In Fig. \ref{fig2c}, the $\mu-P$ relation is given, which satisfies the Vakhitov-Kolokolov (VK) criterion \cite{VK}, i.e., $d\mu/dP < 0$, which is a necessary stability condition for modes supported by self-focusing nonlinearity. Such relation is exactly described by the function of $\mu=-P$, which can be viewed as its unique property. Because the soliton is narrow and most of the energy is confined into one waveguide, we can approximate it as $U=\left(0,\cdots,0,\sqrt{P},0,\cdots,0\right)^{T}$. Therefore, Eq. (\ref{MU}) is expressed as
\begin{eqnarray}
\mu={U^{\dag}(C+V)U\over P}={U^{\dag}VU\over P}={-P^{2}\over P}=-P. \label{MU-P}
\end{eqnarray}
This analysis also explains why such relation is independent of the change of $c_{0}$.

\begin{figure}[tbp]
\centering%
\subfigure[] {\label{fig3a}
\includegraphics[scale=0.25]{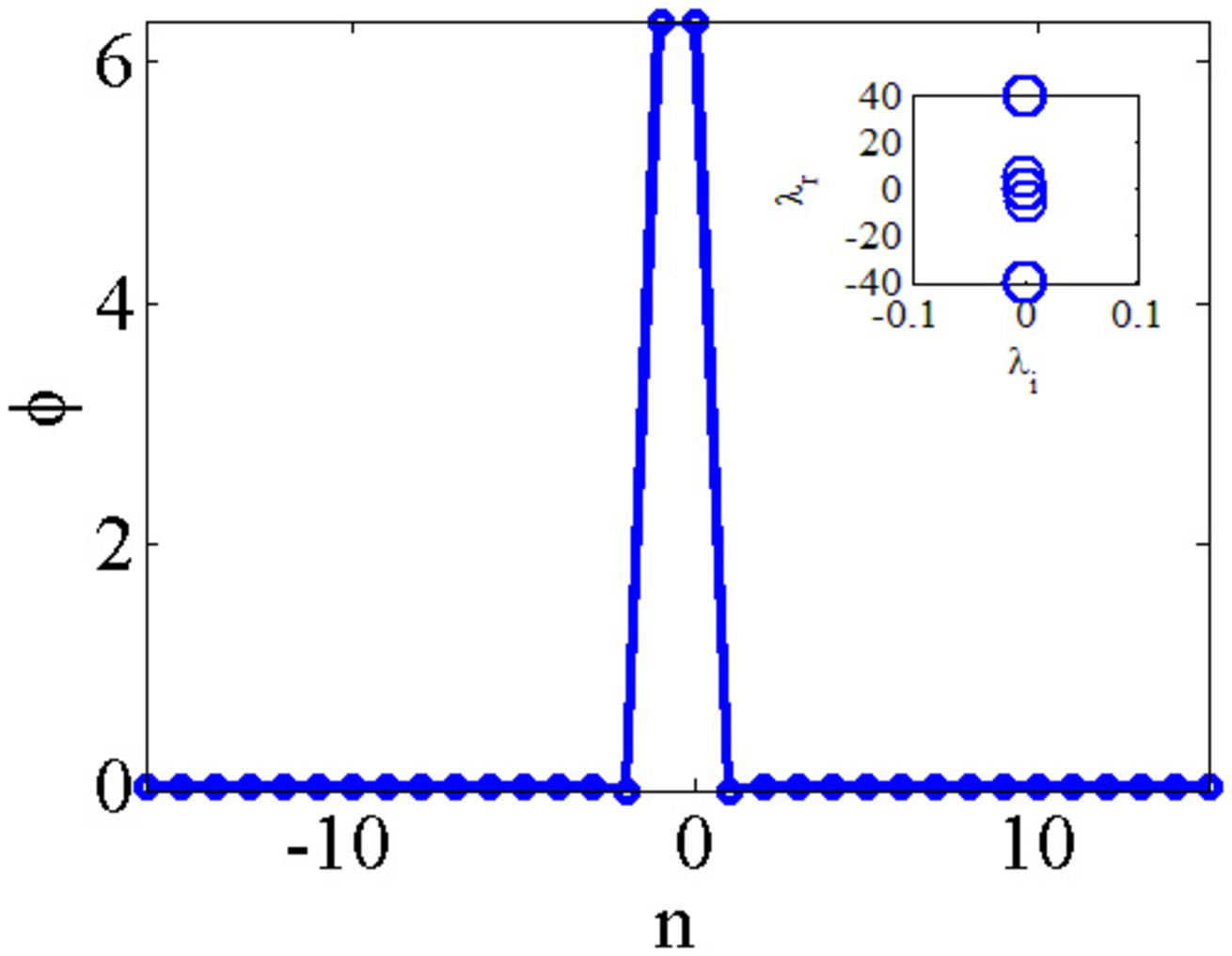}}%
\subfigure[] {\label{fig3b}
\includegraphics[scale=0.25]{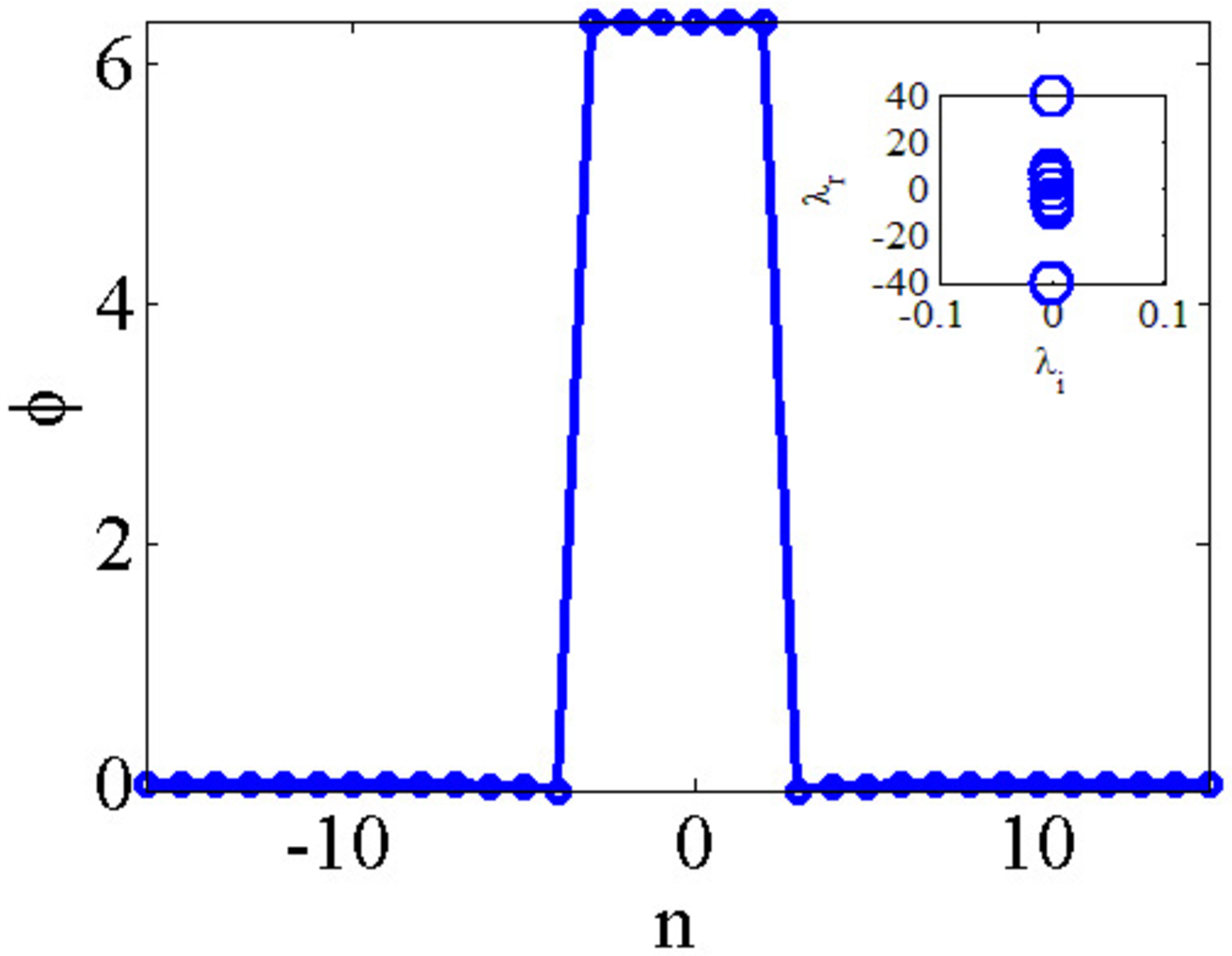}}
\subfigure[] {\label{fig3c}
\includegraphics[scale=0.25]{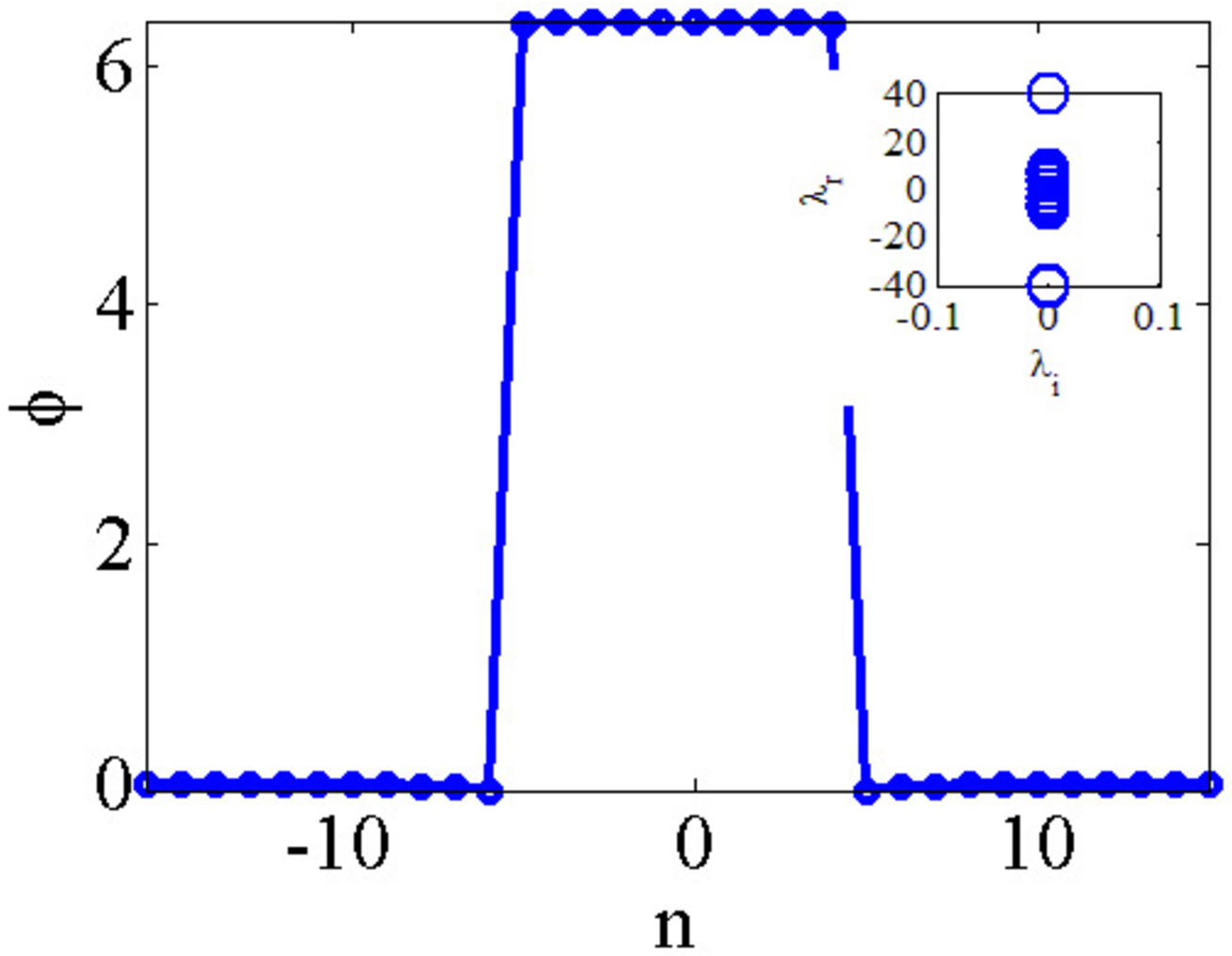}}
\subfigure[] {\label{fig3d}
\includegraphics[scale=0.25]{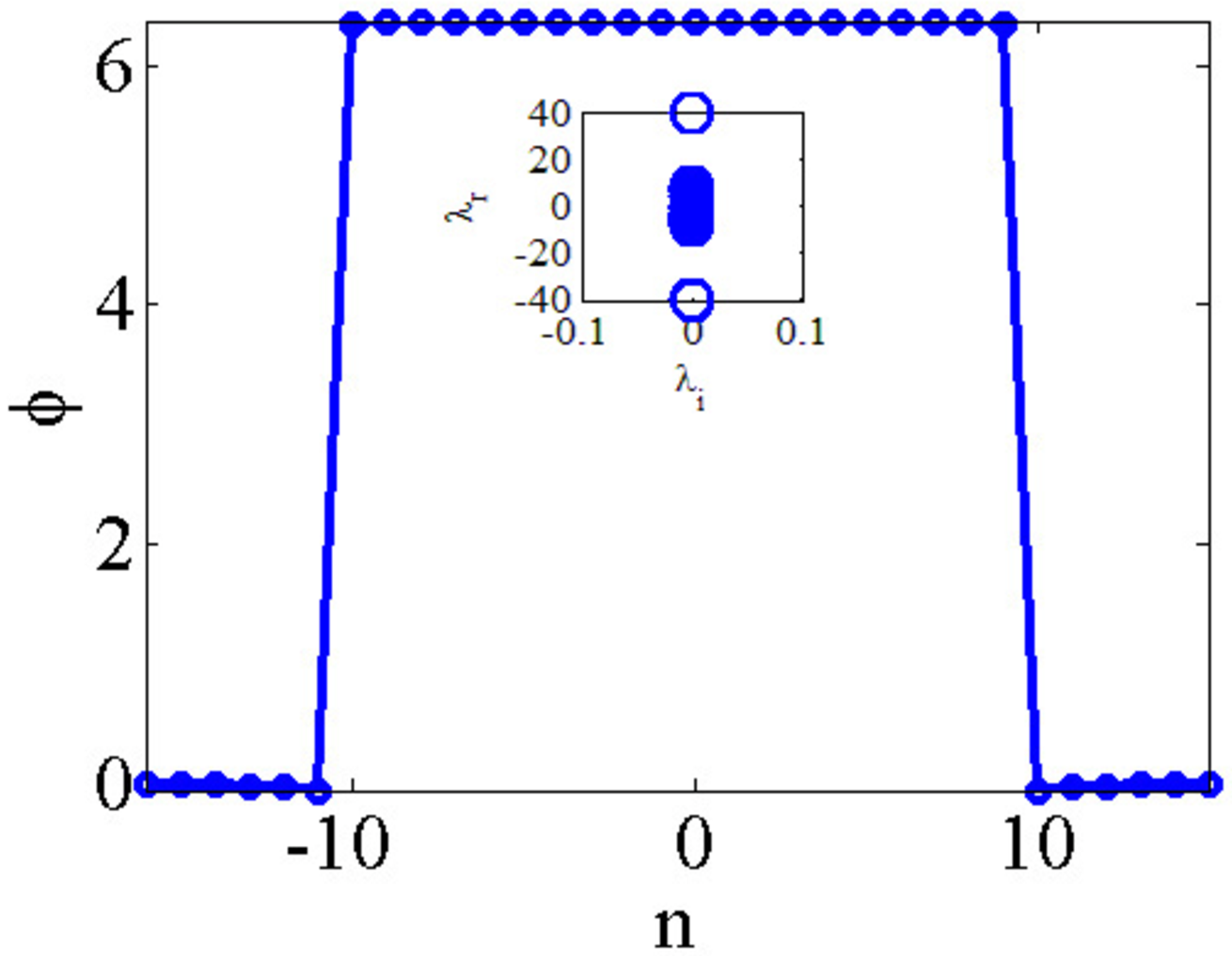}}
\subfigure[] {
\includegraphics[scale=0.25]{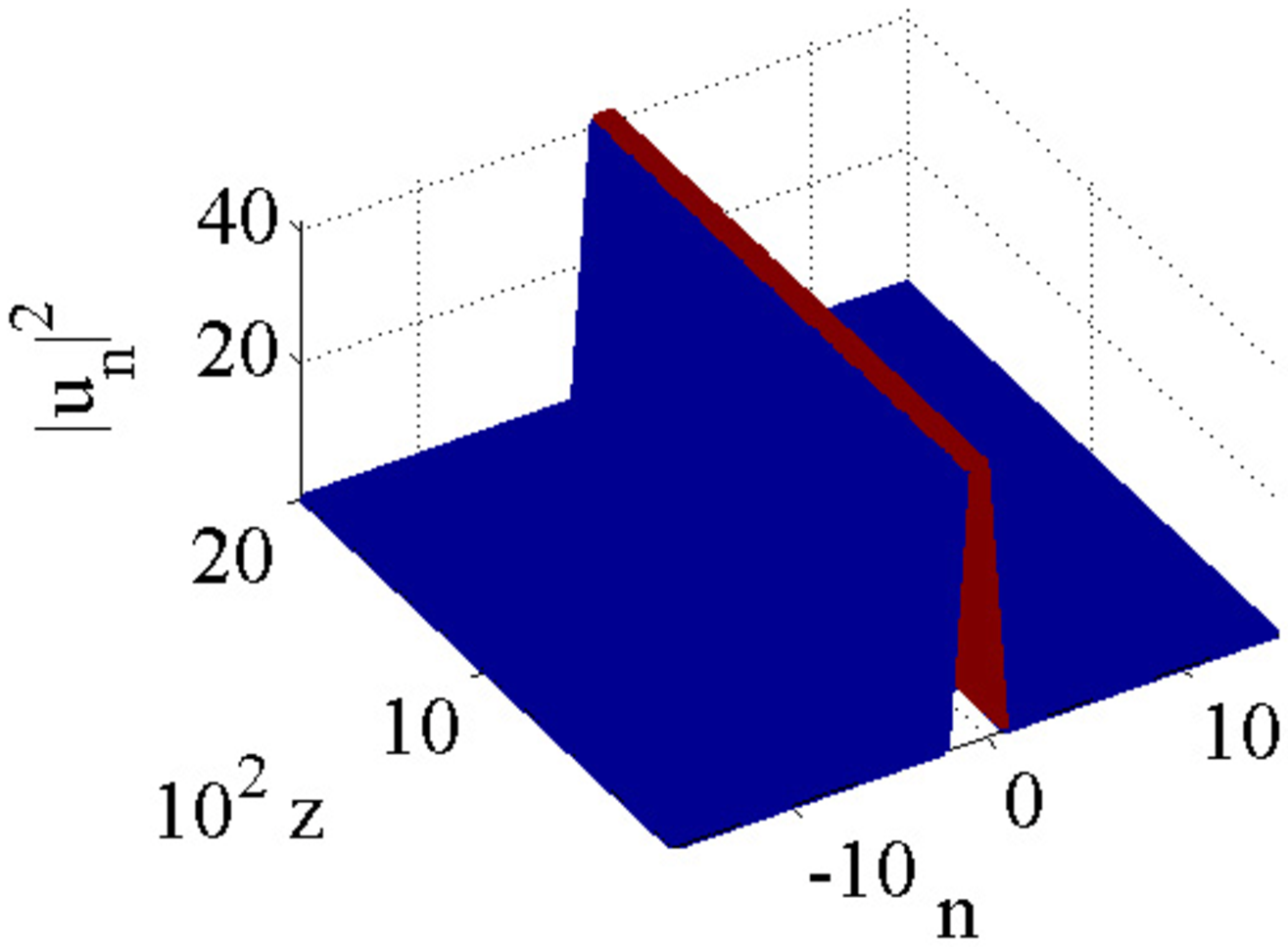}}
\subfigure[] {
\includegraphics[scale=0.25]{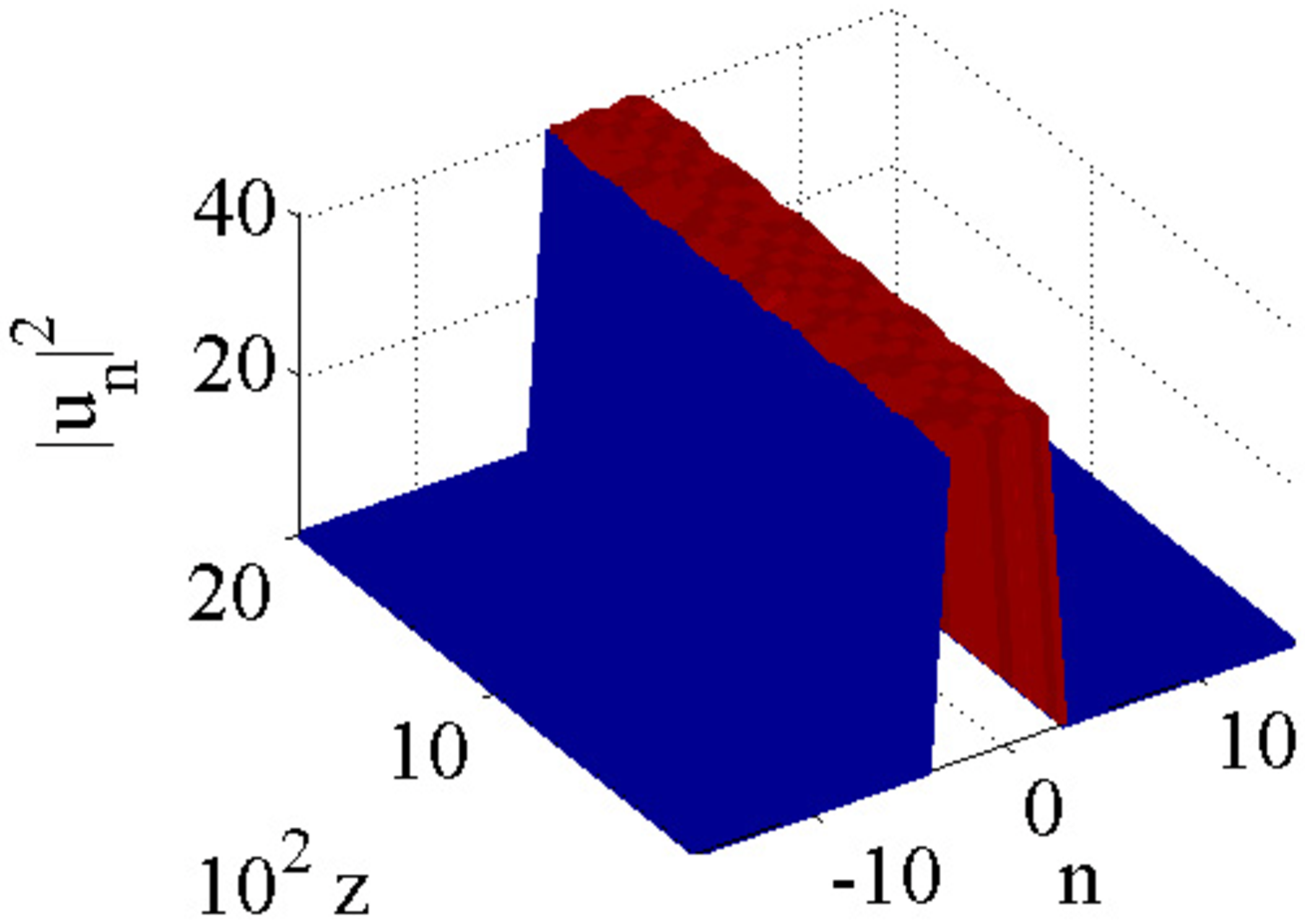}}
\subfigure[] {
\includegraphics[scale=0.25]{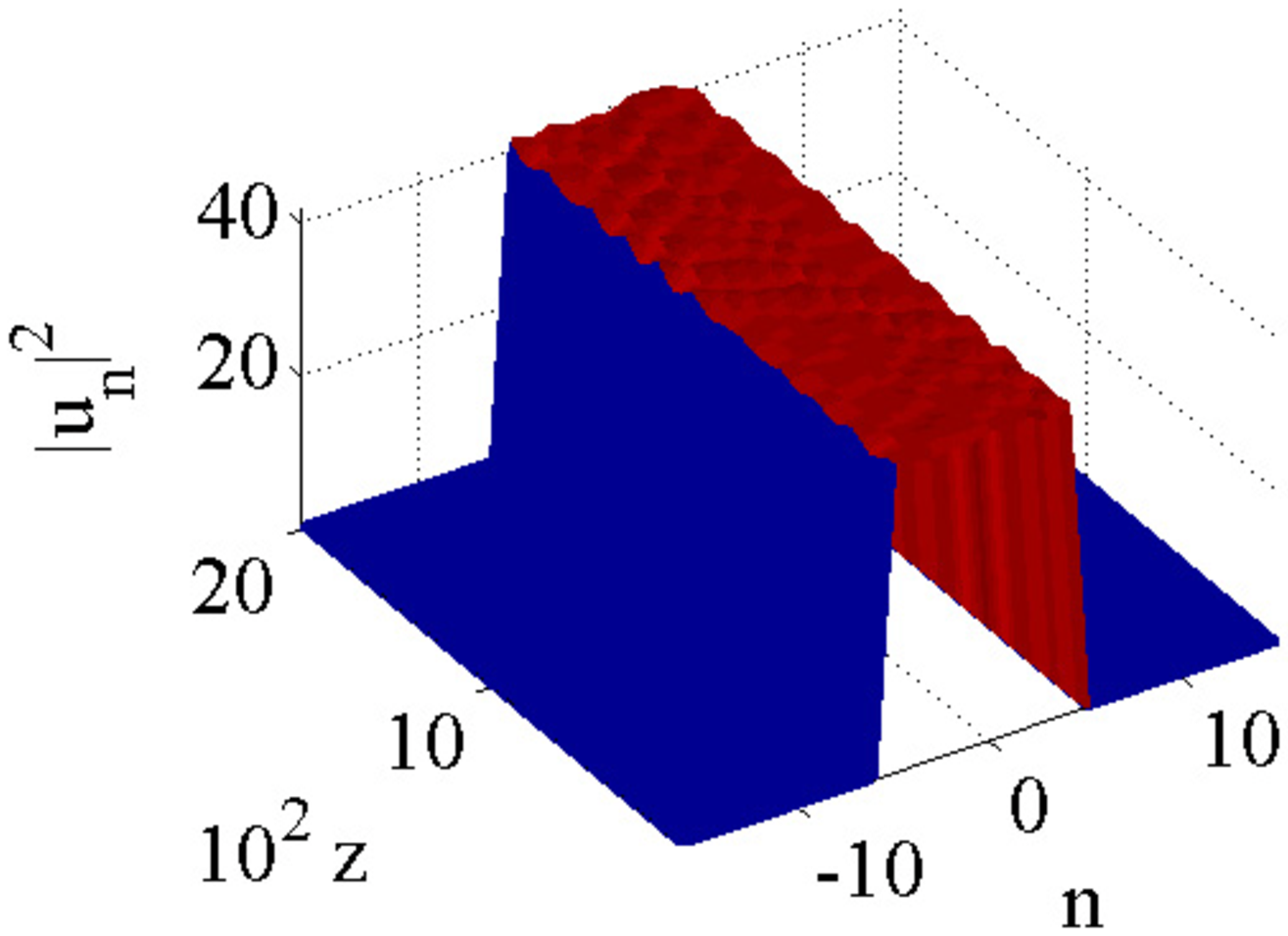}}
\subfigure[] {
\includegraphics[scale=0.25]{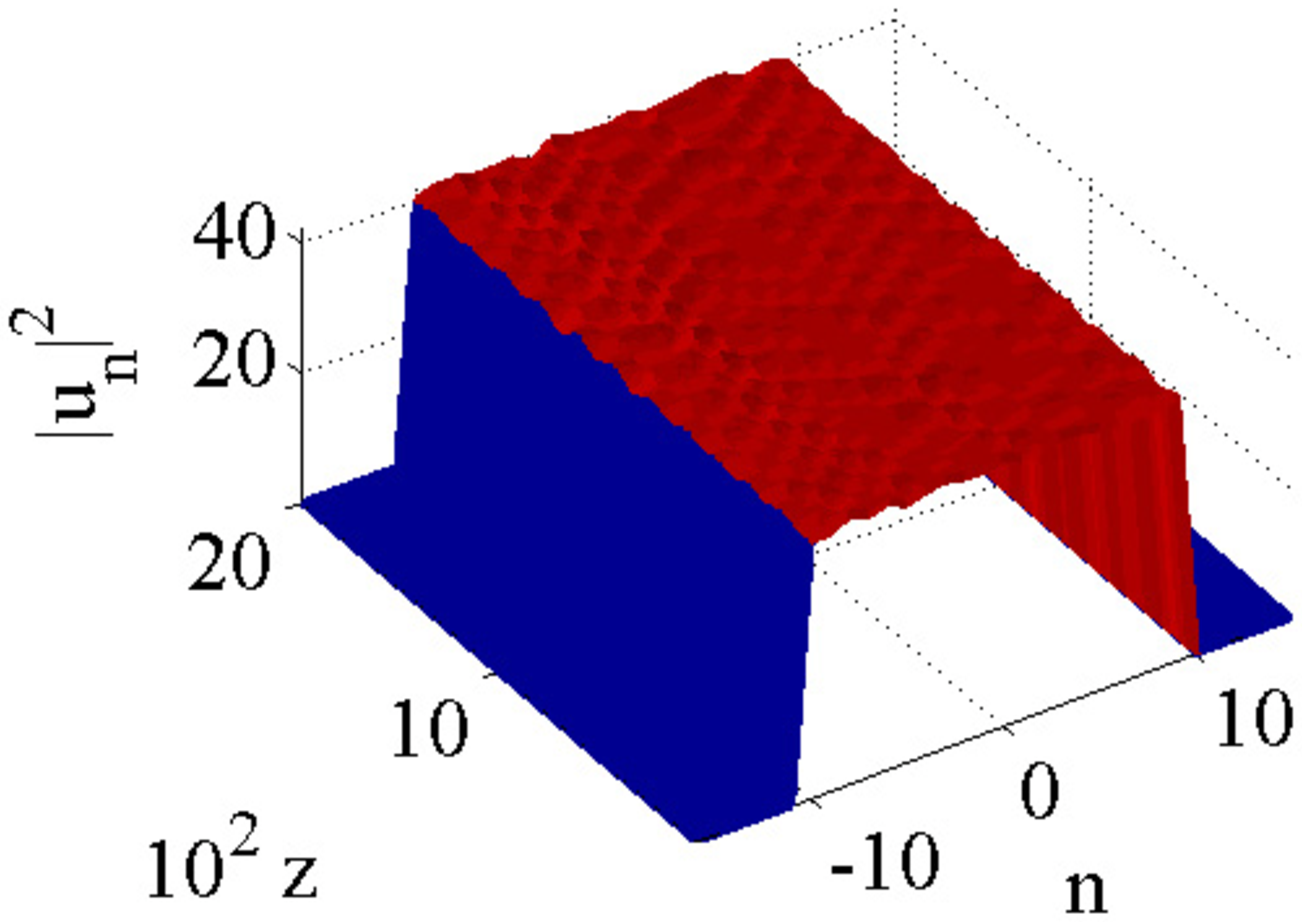}}
\caption{(Color online) Typical example of 2-point ($P=80.4$) (a), 6-point ($P=242.6$) (b), 10-point ($P=404.9$) (c) and 20-point ($P=810.7$) (d) flat-top solitons at $\mu=40$ and $\gamma=1$, insert maps are growth rate of soliton. (e)$\sim$(h) Evolutions of soliton solution in panel (a)$\sim$(d), respectively, which prove such solitons are also stable. }\label{flattop}
\end{figure}

\begin{figure}[tbp]
\centering%
\subfigure[] {\label{fig4c}
\includegraphics[scale=0.5]{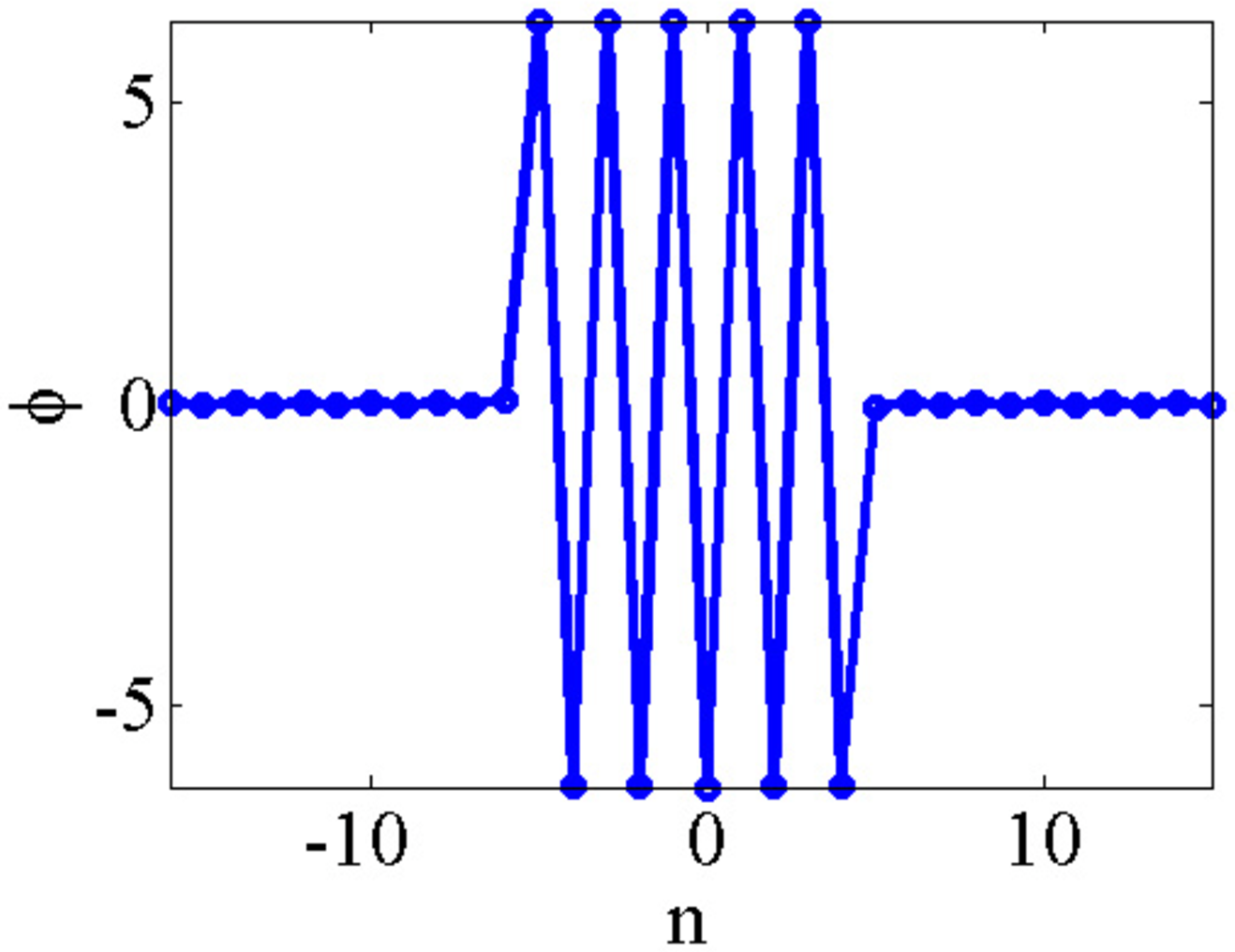}}
\subfigure[] {\label{fig4d}
\includegraphics[scale=0.5]{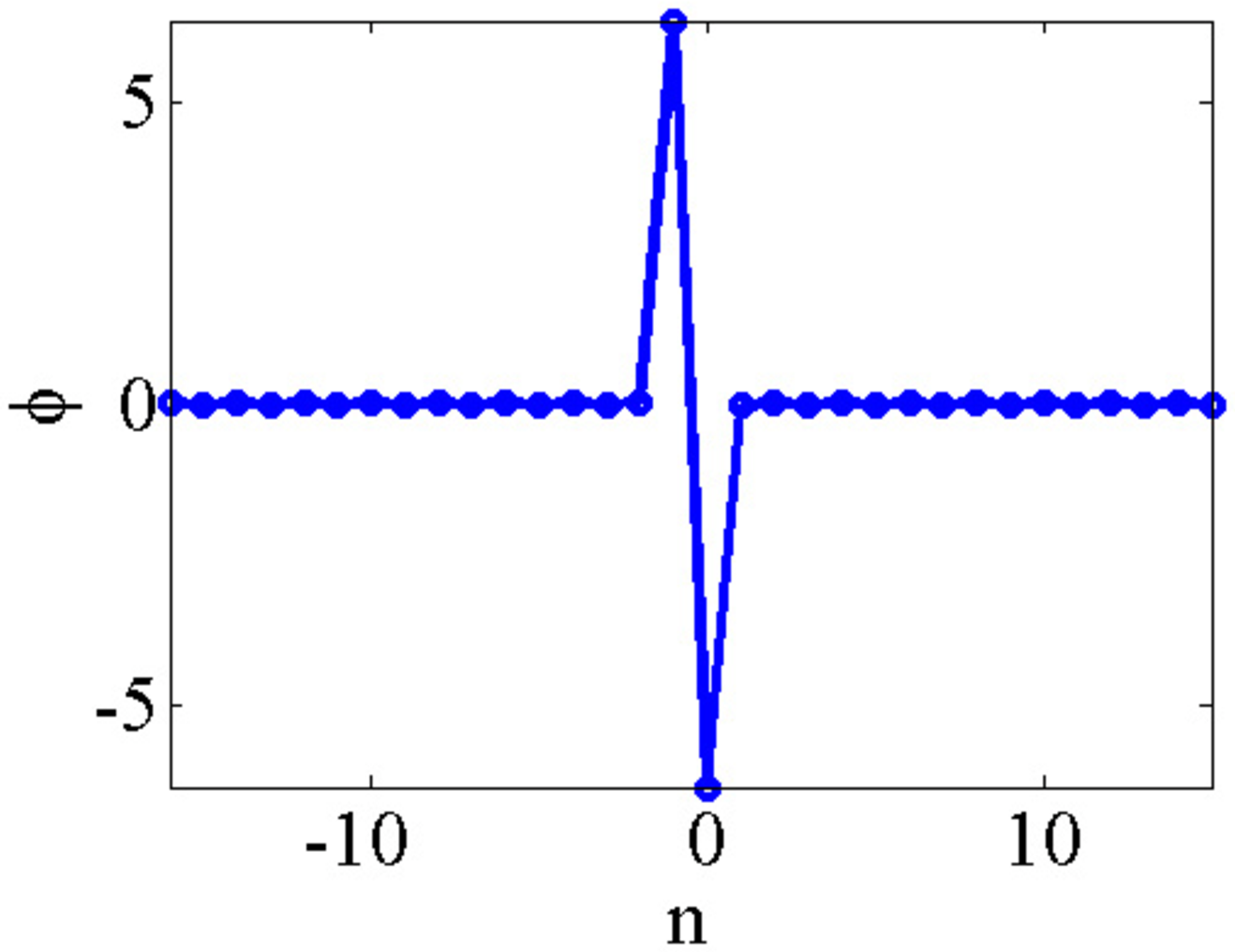}}\\
\subfigure[] {}
\includegraphics[scale=0.5]{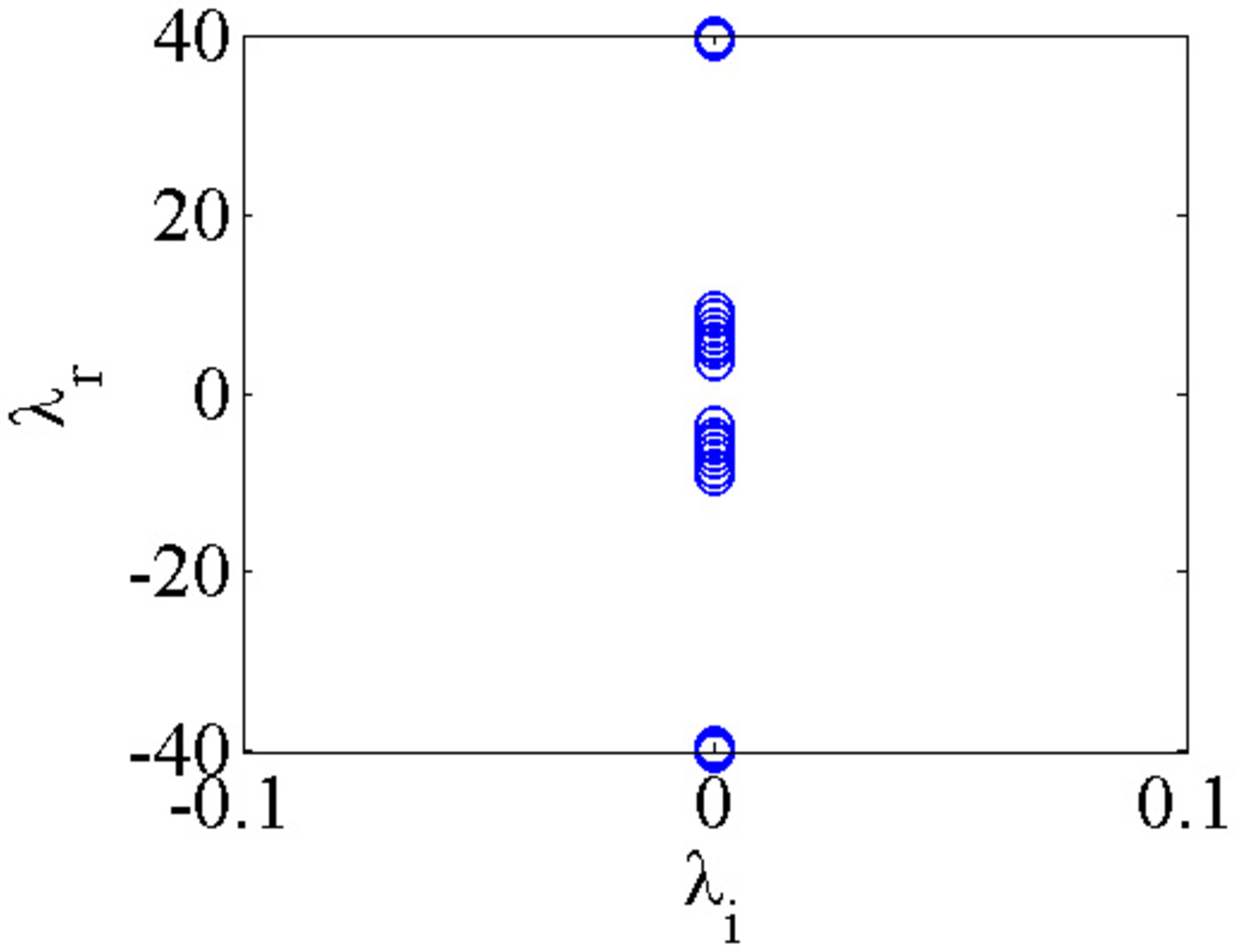}
\subfigure[] {}
\includegraphics[scale=0.5]{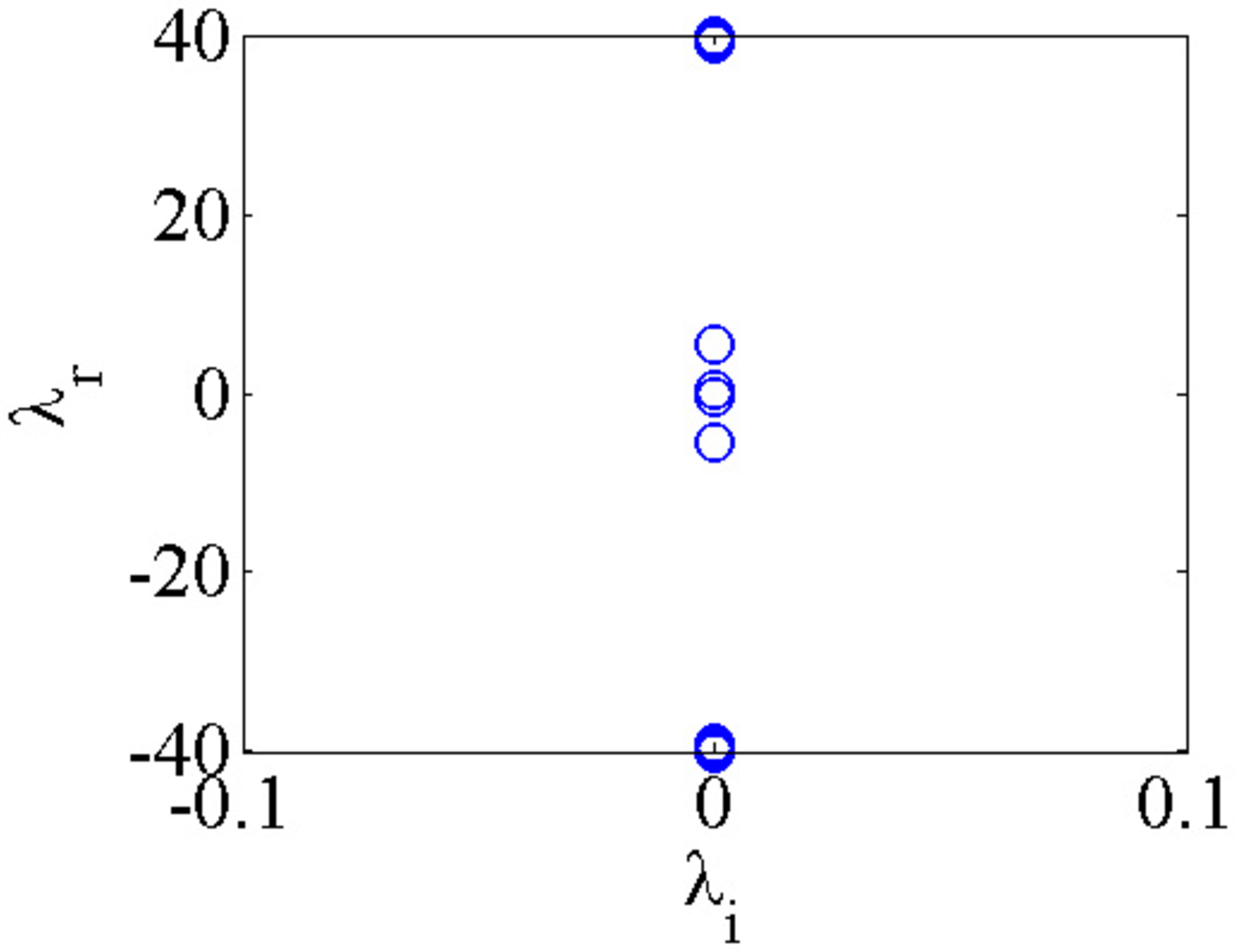}
\caption{(Color online)(a) Stagger mode of 10-point flat-top soliton in self-focusing ($\gamma=-1$) nonlinearity with $\mu=-40$. (b) A dipole soliton staggered from solution in Fig. \ref{fig3a} in the case of self-focusing ($\gamma=-1$). (c)$\sim$(d) Growth rates of soliton in panel (a) and (b), respectively, which prove such solitons are stable.}\label{flattopstabilty}
\end{figure}

To study the excited state soliton, the Newton method is applied \cite{Jianke2}. In the case of self-defocusing ($\gamma=1$), stable flat-top solitons, whose widths (i.e., the number of waveguides occupied by the flat-top soliton) can be controlled by total power, are found. Typical examples are shown in Figs. \ref{flattop}. Although flat-top modes in the discrete system with only the nearest neighboring interaction considered are observed, they are not definitely stable, decaying eventually after propagating for long distances \cite{Darmanyan}.

Fig. \ref{flattop} shows that, for a fixed value of $\mu$, forming different flat-top solitons requires different values of $P$. The relation between $P$ and the number of points is shown in Fig. \ref{fig2d}. The relation of $P-N_{p}$ (where $N_{p}$ is the number of waveguides occupied by the flat-top soliton) can be exactly described by $P=\mu N_{p}$. This relationship is is analyzed as follows: we assume that the $N_{p}$-point flat-top soliton with total power $P$ is described by $U=\left(0,\cdots,0,\sqrt{P_{1}},\cdots,\sqrt{P_{N_{p}}},0,\cdots,0\right)^{T}$, where $P_{1}=P_{2}=\cdots=P_{N_{p}}\equiv P/N_{p}$. Eq. (\ref{MU}) is then written as
\begin{eqnarray}
\mu={U^{\dag}(C+V)U\over P}={U^{\dag}VU\over P}={\sum^{N_{p}}_{i=1}\left(P\over N_{p}\right)^{2}\over P}\equiv {P\over N_{p}}. \label{P-Np}
\end{eqnarray}
The above analysis also demonstrates an anti-VK criterion, i.e., $d\mu/dP >0$, which, as argued by \cite{Sakaguchi1}, may also play the role of a necessary stability condition of solitons in self-defocusing nonlinearity.

By stagger operation, the flat-top mode in the self-defocusing ($\gamma=1$) can stagger to the self-focusing ($\gamma=-1$) case [Fig. \ref{fig4c}]. The stable two-point flat-top soliton in Fig. \ref{fig3a} implies that the system can support double-monopole (even mode) solitons in the self-defocusing case. After the stagger operation, the double-monopole soliton can be converted to a dipole mode, indicating that the system can support stable dipole solitons in the self-focusing case. Figure \ref{fig4d} gives an example of a stable dipole soliton staggered from the solution in Fig. \ref{fig3a}. Further numerical simulations shows that these excited state solitons are also very pinned and cannot be kicked across the waveguide array.

To study of asymptotic properties of the system, we extend Eq.(4) with a normalized decay length $d$, which describes the effective length of the nonlocal effect in the waveguides. Then Eq. (4) can be rewritten as:\\
\begin{eqnarray}
C_{mn}=
\begin{cases}
c_{0}\exp(-j/d) & (j\neq -1) \\
0 & (j=-1).
\end{cases} \label{C_elementnew}
\end{eqnarray}
In Eq. (13), we redefine $j=|m-n|-1$ to make sure the system go back to the nearest-neighbor-coupled model when $d<<1$. In Fig. 6(a), we depict the threshold of power(norm), i.e., $P_{cr}$, above which stable soliton solution can be built, as a function of $d$. In this panel, a minimum threshold value is found approximately at $d=0.42$. On the left and right side of this point, $P_{cr}$ decreases and increases with $b$ increasing, respectively, which indicates the system transits from local to nonlocal effect around this point. Soliton solution with large value of $d$ are also studied. In Fig. 6(b), we plot a typical example at $d=5$. Compare with Fig. 2(a), we can see that soliton with larger value of $d$ gains a wider nonzero tails than the smaller ones.\\
\begin{figure}[tbp]
\centering%
\subfigure[] {\label{}
\includegraphics[scale=0.7]{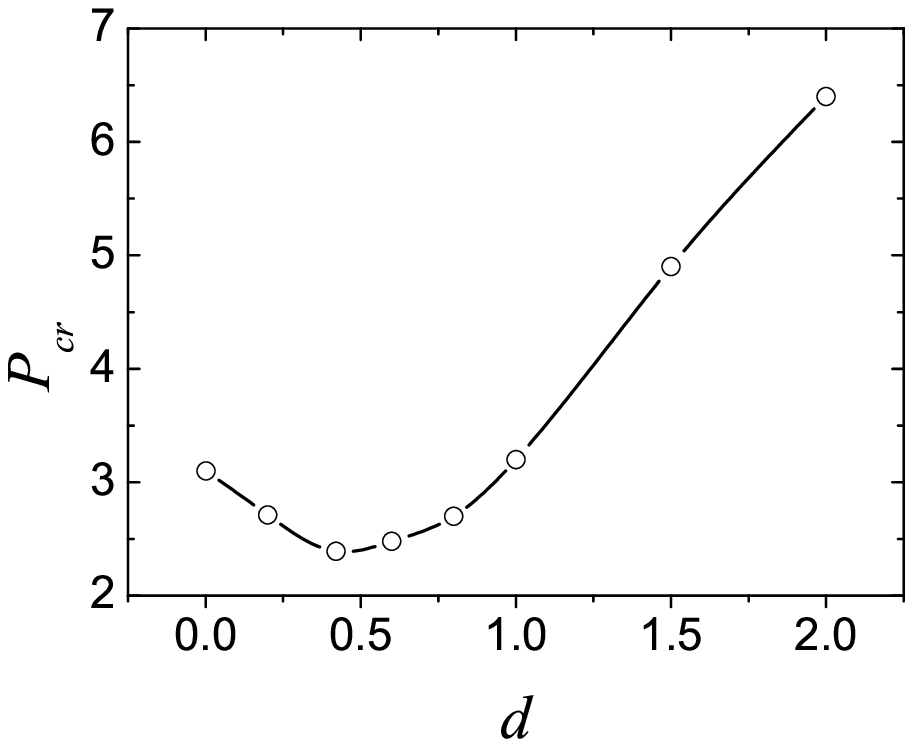}}%
\subfigure[] {\label{}
\includegraphics[scale=0.7]{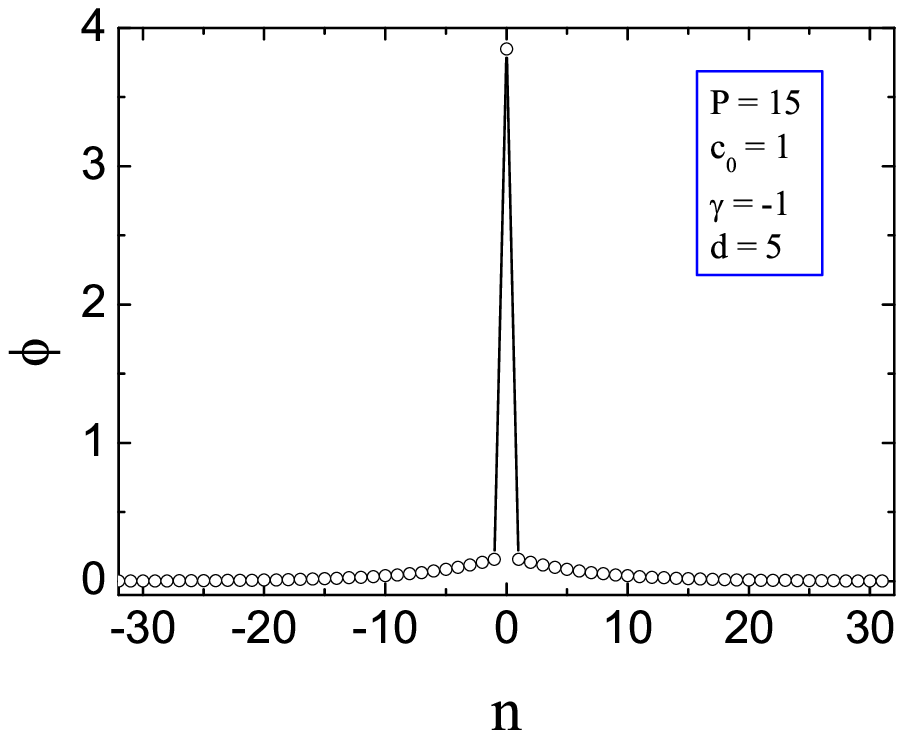}}
\caption{(Color online).(a) $P_{cr}$ as a function of $d$. (b) Fundamental soliton solution in self-attractive/focusing case ($\gamma=-1$) with $c_{0}=1$, $P=15$ and $d=5$ }\label{}
\end{figure}

\section{Conclusion}
This study aims to examine DSs, including their linear long-range interactions. This system is described by the discrete nonlinear Schrodinger equation with all off-diagonal elements in the linearly coupled matrix filled by non-zero interaction terms. The stable fundamental DSs in this system, formed above power threshold values, have narrow widths and occupy almost only one waveguide. These solitons are pinned and cannot be kicked across the waveguides.  For excited-state DSs, stable dipole and double-monopole solitons, which occupy two waveguides, are found in self-focusing and -defocusing nonlinearities, respectively. The stable flat-top soliton and its stagger mode with an arbitrary number of occupied waveguides, which can be tuned by the total power, are observed in these circumstances as well. Because the family of solitons in the system can exhibit abundant digital properties, which may be termed digital discrete solitons, it can give rise to additional data processing applications and have potential for the fabrication of digital optical devices in all-optical networks.

\begin{acknowledgments}
The authors appreciate the useful discussion from Prof. B. A. Malomed and especially thank Prof. Chongjun Jin for his time and efforts on this work and allowing to conduct the results in his labs. This work is supported by the National Natural Science Foundation of China (Grant Nos.11104083, 11204089, 61172011) and the Guangdong Natural Science Foundation (grant No. 10151064201000006). We would like to express our gratitude to the Modern Educational Technology Center of South China Agricultural University for giving us access to its computing facility.
\end{acknowledgments}

%

\bibliographystyle{plain}
\bibliography{apssamp}

\end{document}